\documentstyle[preprint,aps]{revtex}

\begin{document}

\draft

\title{Constraints on  Light Pseudoscalars Implied by Tests
of the Gravitational Inverse-Square Law}

\author{Ephraim Fischbach$^{(1)}$ and Dennis E. Krause$^{(2,1)}$}

\address{$^{(1)}$Physics Department, Purdue University, West Lafayette, IN 
47907-1396}

\address{$^{(2)}$Physics Department, Wabash College, Crawfordsville, IN 
47933-0352}

\date{\today}

\maketitle

\begin{abstract}

The exchange of light pseudoscalars between fermions leads to a
spin-independent potential in order $g^{4}$, where $g$ is the Yukawa
pseudoscalar-fermion coupling constant.  This potential gives rise to
detectable violations of both the weak equivalence principle (WEP) and the
gravitational inverse-square law (ISL), even if $g$ is quite small.  
We show that
when previously derived WEP constraints are combined with those arising
from ISL tests, a direct experimental limit on the
Yukawa coupling of light pseudoscalars to neutrons can be inferred for the
first time ($g_{n}^{2}/4\pi \lesssim 1.6 \times 10^{-7}$), along with a  new
(and significantly improved) limit on the coupling of light pseudoscalars to
protons.

\end{abstract}

\pacs{04.80.Ce, 14.80.-j}

\pagebreak

		In a previous paper \cite{Fischbach 1999}, it was shown that laboratory
bounds on the Yukawa couplings of light pseudoscalars to protons and
neutrons could be significantly improved by using the results from recent
weak equivalence principle (WEP) experiments 
\cite{Gundlach}.  These experiments are
sensitive to the spin-independent long-range forces that arise in order
$g^{4}$ from two-pseudoscalar exchange \cite{Fischbach 1999,Drell,Ferrer},
where
$g$ is defined by the coupling
\begin{equation}
 {\cal L} (x) = ig \bar{\psi}(x)\gamma_5 \psi(x) \phi(x).
\label{L}
\end{equation}
Here $\phi(x)$ is the field operator for a pseudoscalar of mass $m$, and
$\psi(x)$ denotes either a proton ($p$), electron ($e$), or neutron ($n$) of
mass $M_{p}$, $M_{e}$, or $M_{n}$ respectively.  For each pair of interacting
particles, ${\cal L}(x)$ leads to a potential
$V^{(4)}(r)$ in order $g^{4}$ which in the $m \rightarrow 0$ limit is given
by \cite{Drell,Ferrer}

\begin{equation}
 V^{(4)}_{ab} (r) = -\frac{g_{a}^2g_{b}^{2}}{64\pi^3M_{a}M_{b}}\,
\frac{1}{r^3},
\label{V_ps 4}
\end{equation}
where $a$ and $b$ may each denote $p$, $e$, or $n$.  The object of the 
present paper is to demonstrate that already-existing data
from tests of the gravitational
inverse-square law (ISL) \cite{Spero,Fischbach
book} provide new stringent  constraints on $g_{p}^{2}$ and
$g_{n}^{2}$.  When combined with the constraints implied by
Eq.~(\ref{Delta a}) below and the data from the WEP test in
Ref.~\cite{Gundlach}, the ISL  data lead to the first direct
experimental bound on the pseudoscalar-neutron coupling constant
$g_{n}^{2}$, and to a significantly improved bound on $g_{p}^{2}$ (see
Eq.~(\ref{absolute limits}) below).

     Leaving aside for the
moment the contribution from electrons, it was shown in Ref.
\cite{Fischbach 1999} that $V^{(4)}_{ab}$ leads to an acceleration difference
$\Delta\vec{a}_{2-2'}$ of macroscopic
test objects 2 and $2'$ in the presence of a common source $M_{1}$.  If these
have  masses $M_{2}$ and
$M_{2'}$, and contain $Z_{2}$ ($N_{2}$) protons (neutrons), and $Z_{2'}$
($N_{2'}$) protons (neutrons) respectively, then 
\begin{equation}
\Delta\vec{a}_{2-2'} = \vec{\cal F}(\vec{r})
     \left(\frac{M_{1}}{m_{H}^{2}}\right)
   \left[ g_{p}^{2}\left(\frac{Z_{1}}{\mu_{1}}\right) +
         g_{n}^{2}\left(\frac{N_{1}}{\mu_{1}}\right)\right]
  \left[ g_{p}^{2}\,\Delta\!\left(\frac{Z}{\mu}\right)_{2-2'} +
         g_{n}^{2}\,\Delta\!\left(\frac{N}{\mu}\right)_{2-2'}\right].
\label{Delta a}
\end{equation}
In Eq.~(\ref{Delta a}), $\vec{\cal F}(\vec{r})$ is the integral over the
mass distribution of the source \cite{Fischbach 1999},
$\mu_{i} = M_{i}/m_{\rm H}$, $m_{\rm H}~=~m(_{1}{\rm H}^{1})$, 
$M_{n} \simeq M_{p} \equiv M$,
and $\Delta(Z/\mu)_{2-2'} = Z_{2}/\mu_{2} - Z_{2'}/\mu_{2'}$, etc.,  
\cite{Fischbach book}.  Since all the parameters appearing in
Eq.~(\ref{Delta a}) are known, except for the pseudoscalar couplings
$g_{p}^{2}$ and $g_{n}^{2}$ (to protons and neutrons respectively), an
experimental determination of $\Delta\vec{a}_{2-2'}$ leads to a constraint
on $g_{p}^{2}$ and $g_{n}^{2}$.

As noted in Ref.~\cite{Fischbach 1999}, however, the right-hand side of
Eq.~(\ref{Delta a}) vanishes whenever $g_{p}^{2}$ and $g_{n}^{2}$ satisfy
\begin{equation}
\frac{g_{p}^{2}}{g_{n}^{2}} =
      - \frac{\Delta(N/\mu)_{2-2'}}{\Delta(Z/\mu)_{2-2'}},
\label{no limit condition}
\end{equation}
in which case $g_{p}^{2}$ and $g_{n}^{2}$ can be arbitrarily large and
still be compatible with any experimental bound on
$\Delta\vec{a}_{2-2'}$.  Since the right-hand side of Eq.~(\ref{no limit
condition}) is close to 1 for most pairs of materials, including those used
in Ref.~\cite{Gundlach}, Eq.~(\ref{no limit condition}) can be satisfied
even when $g_{p}^{2}$ and $g_{n}^{2}$ are each quite large provided
$g_{p}^{2}\simeq g_{n}^{2}$.  This is shown graphically in Fig. 1 which
plots the constraints in the $g_{p}^{2}$-$g_{n}^{2}$ plane that emerge
when Eq.~(\ref{Delta a}) is combined with the experimental limits of
Gundlach, {\em et al.} \cite{Gundlach}.  It is seen that the boundary of
the allowed region is a hyperbola with an asymptote near $g_{p}^{2} =
g_{n}^{2}$, along which no limits on $g_{p}^{2}$ or $ g_{n}^{2}$ can be
inferred.  To circumvent the problem caused by such ``hyperbolic''
constraints, one can combine results from experiments using different
materials, which thus have slightly different asymptotes. 
Alternatively one can 
choose special materials (such as 2 = Li and $2'$ = Ru) for which
Eq.~(\ref{no limit condition}) can never hold \cite{Fischbach 1999}, and
which thus lead to ellipses in the $g_{p}^{2}$-$g_{n}^{2}$ plane.  The
combination of ``elliptical'' and ``hyperbolic'' constraints would then
lead to separate bounds on $g_{p}^{2}$ and $g_{n}^{2}$.   As we now
demonstrate, existing data from ISL tests also provide elliptical constraints
on $g_{p}^{2}$ and $g_{n}^{2}$ and, when combined with earlier WEP results,
lead directly to the bounds quoted in Eq.~(\ref{absolute limits}) below.

Consider the ISL experiment of Spero, {\em et al.}\cite{Spero,Fischbach book} in
which a cylindrical Cu test mass is suspended at the end of a torsion
fiber inside a larger hollow stainless steel cylinder.  It can be
shown that for infinitely long cylinders the Cu test mass will
experience a force from the stainless steel cylinder only if the
underlying interaction is not a pure $1/r^{2}$ force.  When the
small (and calculable) end effects due to finite cylinders are
taken into account, the experiment of Spero, {\em et al.} becomes
a null test for the presence of new non-Newtonian 
inverse-power-law interactions,
such as $V^{(4)}_{ab}(r)$ in Eq.~(\ref{V_ps 4}).  We will use 
two convenient parameterizations of power law potentials between
two particles 1 and 2 \cite{Riley parameterization,Feinberg and Sucher}:
\begin{mathletters}
\label{V_n}
\begin{eqnarray}
V_{n}(r) & = & 
         -\alpha_{n}
   \left(\frac{G_{N}M_{1}M_{2}}{r}\right)
\left(\frac{r_{0}}{r}\right)^{n
- 1}, \\
 & =  & -\Lambda_{n}\left(\frac{B_{1}B_{2}}{r}\right)
\left(\frac{r_{0}}{r}\right)^{n - 1}. 
\end{eqnarray}
\end{mathletters}
Here $G_{N}$ is the Newtonian gravitational constant, $r_{0} =
1$ fm is an arbitrarily chosen length scale, $B_{1}$ and $B_{2}$
are the baryon numbers for bodies 1 and 2 respectively, and
$\Lambda_{n}$ and $\alpha_{n}$ are dimensionless constants
characterizing the strength of the interaction.   When gravity is included,
the total potential energy  between these two point masses is given by
\begin{mathletters}
\label{V_tot}
\begin{eqnarray}
V_{\rm tot}(r) & =  & -\frac{G_{N}M_{1}M_{2}}{r}\left[1 +
\alpha_{n}\left(\frac{r_{0}}{r}\right)^{n - 1}\right], \\
          & = &  -\frac{G_{N}M_{1}M_{2}}{r}
     \left[1 +
\Lambda_{n}\left(\frac{m_{P}}{m_{H}}\right)^{2}
      \left(\frac{B_{1}}{\mu_{1}}\right)
      \left(\frac{B_{2}}{\mu_{2}}\right)
       \left(\frac{r_{0}}{r}\right)^{n - 1}\right],
\end{eqnarray}
\end{mathletters}
where $m_{P} \equiv \sqrt{\hbar c/G_{N}}$ is the Planck mass.
The null results of the ISL test
of Spero, {\em et al.} can then be used to set limits on $\alpha_{n}$ or
$\Lambda_{n}$, after integrating the corresponding $1/r^{n + 1}$ force
laws over the mass distributions of the Cu test mass and the stainless
steel cylinder \cite{Riley}.  The $1\sigma$ limits implied by Spero, {\em
et al.} for $\alpha_{n}$ and $\Lambda_{n}$ are shown in Table I for
several physically relevant values of $n$.  The results in Table I, which
were obtained by direct integration over the mass distributions of the
interacting Cu and stainless steel cylinders, are in excellent agreement
with those obtained previously by Mostepanenko and Sokolov \cite{MS} who used
a phenomenological parameterization of the non-Newtonian interaction to
constrain $\Lambda_{n}$. Although we are specifically concerned with the case
$n = 3$, other values of $n$ are also interesting:
$n = 2$ potentials can arise from 2-scalar exchange, as well as
2-photon exchange \cite{2-photon}, and $n = 5$ characterizes the 2-body
potential from neutrino-antineutrino exchange \cite{neutrino} and the
2-pseudoscalar exchange potential with derivative coupling (which is
applicable to axions) \cite{Ferrer}.  Note, however, that
$n = 1$ is uninteresting since such a potential would not lead to a deviation
from the inverse-square law, but only to a modified value of $G_{N}$ (which
would be difficult to detect).  Table I also presents the $1\sigma$ limits
derived from the experiment of Mitrofanov and Ponomareva (MP) \cite{MP}
which is a test of the ISL over the range 3.8--6.5 mm.  In this experiment a
modified Cavendish apparatus is used to measure the force between a mass A
suspended at one end of a torsion balance, and a second mass B whose
distance from A is varied.  The experiment then compares the
experimental value for the force ratio $F(r_{1})/F(r_{2})$, where
$r_{1} = 3.773(40)$ mm and $r_{2} = 6.473(40)$ mm, to the
calculated ratio expected assuming Newtonian gravity.
We see from Table I that for the case $n = 3$, which is
our concern in this paper, the limits implied by Spero, {\em et al.} are
more stringent than those of MP, although for $n$ = 4, 5, 6 the reverse is
true.

To extract constraints on the pseudoscalar coupling constants $g_{p}^{2}$ and
$g_{n}^{2}$ from Spero, {\em et al.}, we begin by considering the interaction
of two macroscopic objects separated by a distance $r$ that is large
compared to their dimensions.  From Eq.~(\ref{V_ps 4}) the total
2-body interaction energy $V_{12}^{(4)}$  in order $g^{4}$ is given by
\begin{equation}
V^{(4)}_{12}(r)  = 
           -\frac{1}{64\pi^{3}M^{2}}
            \,\frac{1}{r^{3}}
        \left(g_{p}^{2}Z_{1} + g_{n}^{2}N_{1}\right)
            \left(g_{p}^{2}Z_{2} + g_{n}^{2}N_{2}\right),
\label{V4-macro}
\end{equation}
which should be compared with the parameterizations of
Eq.~(\ref{V_n}) for $n = 3$.
For the actual geometry of the Spero experiment one must
integrate over the mass distributions of the inner test mass and
the outer cylinder \cite{Riley}, so that $1/r^{3}$ in Eqs.~(\ref{V4-macro})
and (\ref{V_n}) is replaced by the appropriate average $\langle
1/r^{3}\rangle$.  By combining Eqs.~(\ref{V4-macro}) and
(\ref{V_n}) for $n = 3$ the constraint implied by Spero, {\em et
al.} can be expressed in the form
\begin{equation}
	\left(g_{p}^{2}\frac{Z_{1}}{\mu_{1}} + g_{n}^{2}
           \frac{N_{1}}{\mu_{1}}\right)
	\left(g_{p}^{2}\frac{Z_{2}}{\mu_{2}} + g_{n}^{2}
           \frac{N_{2}}{\mu_{2}}\right)
		= 
           64\pi^{3}G_{N}M^{2}m_{H}^{2}\alpha_{3}r_{0}^{2}.
\label{limit equation}
\end{equation}
Using the $1\sigma$ limits from Spero, {\em et al.}  presented in Table I we
then find
\begin{equation}
  \left(0.469g_{p}^{2} + 0.540g_{n}^{2}\right)
\left(0.460g_{p}^{2} + 0.549g_{n}^{2}\right) \lesssim 3.5 \times
10^{-12}.
\label{g2 limits}
\end{equation}
In Eq.~(\ref{g2 limits}) the expression in the first set of parentheses
arises from the hollow cylinder, where the coefficients of $g_{p}^{2}$ and
$g_{n}^{2}$ are the values of $Z/\mu$ and $N/\mu$, respectively, for
stainless steel.  Similarly, the expression in the second set of
parentheses represents the contributions from the Cu test mass, while the
right hand side of Eq.~(\ref{g2 limits}) is derived from the bound on
$\alpha_{3}r_{0}^{2}$ quoted in Table I.

	We see immediately from Eq.~(\ref{g2 limits}) that the constraint implied
by the ISL experiment of Spero, {\em et al.} \cite{Spero} leads to an
ellipse in the $g_{p}^{2}$-$g_{n}^{2}$ plane, as can be seen in Fig.~1. 
This is, of course, related to the fact that the left side of Eq.~(\ref{g2
limits}) cannot vanish unless both $g_{p}^{2}$ and $g_{n}^{2}$ do.  Figure~1
also exhibits the previously derived WEP constraint \cite{Fischbach 1999}
from the experiment of Gundlach, {\em et al.} \cite{Gundlach}, which gives
rise to a hyperbola in the $g_{p}^{2}$-$g_{n}^{2}$ plane as we have noted
previously.  The significant new feature of Fig.~1 is that the combination of
the hyperbolic WEP constraint and the elliptical ISL constraint lead to upper
bounds on  $g_{p}^{2}$ and $g_{n}^{2}$ separately.  We find from the figure
the following $1\sigma$ limits:
\begin{mathletters}
\label{absolute limits}
\begin{eqnarray}
g_{p}^{2}/4\pi & \lesssim & 1.6 \times 10^{-7},
     \\
g_{n}^{2}/4\pi & \lesssim & 1.6 \times 10^{-7}.
\end{eqnarray}
\end{mathletters}
The result for $g_{n}^{2}$ in Eq.~(\ref{absolute limits}) represents the
first direct laboratory constraint on the Yukawa coupling of pseudoscalars
to neutrons.  We note that the only previous laboratory limit on $g_{n}^{2}$
\cite{Fischbach 1999} was based on an indirect model-dependent argument due
to Daniels and Ni \cite{Daniels} utilizing the spin-dependent results
of Ritter, {\em et al.} \cite{Ritter}.
For $g_{p}^{2}$ there is an earlier result due to Ramsey
\cite{Ramsey}, $g_{p}^{2}/4\pi \lesssim 2.5 \times 10^{-5}$ ($1\sigma$),
which was obtained from a study of the molecular spectrum of H$_{2}$.  As we
see from Eq.~(\ref{absolute limits}), the limit implied by  combining the ISL
and WEP results improves the Ramsey limit by more than two orders of
magnitude.

	We can also extract from Fig.~1 constraints on $g_{p}^{2}$ and $g_{n}^{2}$
in special cases of interest.  For a universal coupling to baryon number
$g_{p}^{2} = g_{n}^{2}$, and we find (at the $1\sigma$ level)
\begin{equation}
g_{p,n}^{2}/4\pi \lesssim 1.5 \times 10^{-7}.
\label{baryon limit}
\end{equation}
The similarity of the results of Eqs.~(\ref{absolute limits}) and
(\ref{baryon limit}) arises because the largest allowed 
values of $g_{p}^{2}$ and $g_{n}^{2}$  lie near the line 
$g_{p}^{2} = g_{n}^{2}$, as can be seen in Fig. 1.
Two other results of interest are the limiting cases $g_{p}^{2}\gg
g_{n}^{2}$ and $g_{n}^{2} \gg g_{p}^{2}$.  We can see from Fig. 1 that these
are determined by the WEP results of Gundlach, {\em et al.}, and hence can
be taken over from our previous analysis \cite{Fischbach 1999}:
\begin{mathletters}
\label{other limits}
\begin{eqnarray}
g_{p}^{2}/4\pi & \lesssim & 9 \times 10^{-8}, \,\,\,(g_{p}^{2}\gg
g_{n}^{2}), 
\\
g_{n}^{2}/4\pi & \lesssim & 7 \times 10^{-8}, \,\,\,(g_{n}^{2}\gg
g_{p}^{2}).
\end{eqnarray}
\end{mathletters}

		Although we have focused thus far on the pseudoscalar couplings to protons
and neutrons, it is straightforward to show that the contributions from
electrons can be incorporated via the substitution $g_{p}^{2} \rightarrow
g_{c}^{2} \equiv g_{p}^{2} + (M/M_{e})g_{e}^{2}$, where $g_{e}$ is the
pseudoscalar-electron coupling constant.  This follows from 
Eq.~(\ref{V_ps 4}) by noting that the contributions from electrons are
enhanced by a factor $M/M_{e}$ relative to those from protons and neutrons. 
The limits on $g_{p}^{2}/4\pi$ in Eqs.~(\ref{limit equation})--(\ref{other
limits}) can then be taken over immediately for $g_{c}^{2}/4\pi$, and these
lead to constraints on $g_{e}^{2}$, at least in principle.  In practice,
however, existing limits on $g_{e}^{2}/4\pi$ obtained from spin-dependent
experiments \cite{Gillies} are more stringent, $g_{e}^{2}/4\pi
\lesssim 10^{-16}$.  It follows that despite the enhancement arising from
the factor
$M/M_{e}$, the contribution from the term in $g_{c}^{2}/4\pi$ proportional
to $g_{e}^{2}$ is at most of order $10^{-13}$.  Thus, the bounds on
$g_{e}^{2}/4\pi$ implied by Eqs.~(\ref{limit equation})--(\ref{other
limits}) are in fact bounds on $g_{p}^{2}/4\pi$, and the prospects for
constraining $g_{e}^{2}$ via $V^{(4)}_{ab}$ seem quite remote at present.

	The limits on $g_{p}^{2}$ and $g_{n}^{2}$ in
Eqs.~(\ref{g2 limits})--(\ref{other limits}) are the most restrictive direct
laboratory constraints currently available.  Although astrophysical
arguments based on stellar cooling calculations are more stringent
\cite{astro limits}, typically
$g_{p,n}^{2}/4\pi \lesssim 10^{-21}$, they are necessarily more model
dependent.  For derivative-coupled pseudoscalars such as axions, there is at
present no viable alternative to astrophysical bounds, since those arising
from existing laboratory experiments (corresponding to $n = 5$ in Table I)
are too weak to be of use.  However, by adapting the present formalism
future laboratory experiments carried out over shorter distance scales may
give rise to useful bounds on axions, as we will discuss in more detail
elsewhere.

  In summary, we have shown that the limits implied by the ISL experiment of
Spero, {\em et al.} \cite{Spero}, complement those previously derived from
the WEP experiment of Gundlach, {\em et al.} \cite{Gundlach}, and together
allow the pseudoscalar-neutron coupling constant $g_{n}^{2}$ to be directly
determined for the first time.  In addition, the combination of these two
experiments leads to a new bound on the pseudoscalar-proton coupling
constant $g_{p}^{2}$, which improves on the earlier Ramsey limit
\cite{Ramsey} by more than two orders of magnitude.  As was noted in
Ref.~\cite{Fischbach 1999}, the Gundlach results lead to hyperbolic
constraints on $g_{p}^{2}$ and $g_{n}^{2}$ which admit  the possibility
that each of these could be quite large, provided that $g_{p}^{2} \simeq
g_{n}^{2}$.  Absolute bounds on these constants could be obtained,
however, by using special materials such as Li and Ru.  What we have shown
here is that {\em already existing} data from the ISL experiment of Spero,
{\em et al.} provide the needed elliptical constraints, in effect playing
the same role that a WEP experiment utilizing Li and Ru would.  The fact
that the WEP and ISL experiments complement each other in this way raises
the possibility of a new generation of WEP and ISL experiments whose
results, when combined, would lead to even more stringent constraints on
$g_{p}^{2}$ and $g_{n}^{2}$.  Although the laboratory constraints may not be
as restrictive as those implied by astrophysical limits \cite{astro limits},
they are completely model-independent, and furthermore allow $g_{p}^{2}$ and
$g_{n}^{2}$ to be separately determined.

  The authors are deeply indebted to Riley Newman for providing the
limits on $\alpha_{n}r_{0}^{n - 1}$ quoted in Table I, and for numerous
helpful discussions.  This work was supported in part by the U.S. Department
of Energy under Contract No. DE-AC 02-76ER01428.

\begin{table}
\squeezetable
\begin{tabular}{cdddd}
& \multicolumn{2}{c}{Spero, {\em et al.}} 
& \multicolumn{2}{c}{Mitrofanov and Ponomareva} \\
$n$  
& $\alpha_{n}r_{0}^{n - 1}$ & $\Lambda_{n}$ 
& $\alpha_{n}r_{0}^{n - 1}$ & $\Lambda_{n}$  \\
\tableline
2 & 1.3 $\times 10^{-6}$ m & 7.7 $\times 10^{-30}$ 
& 6.8 $\times 10^{-5}$ m & 4.0 $\times 10^{-28}$ 
\\ 
3 & 1.3 $\times 10^{-8}$ m$^{2}$ & 7.7 $\times 10^{-17}$
& 8.1 $\times 10^{-8}$ m$^{2}$ & 4.7 $\times 10^{-16}$
\\ 
4 & 1.7 $\times 10^{-10}$ m$^{3}$ & 9.9 $\times 10^{-4}$ 
 & 1.3 $\times 10^{-10}$ m$^{3}$ & 7.5 $\times 10^{-4}$
\\ 
5 & 2.3 $\times 10^{-12}$ m$^{4}$ & 1.4 $\times 10^{10}$
& 2.1 $\times 10^{-13}$ m$^{4}$ & 1.2 $\times 10^{9}$
\\ 
6 & 3.2 $\times 10^{-14}$ m$^{5}$ & 1.8 $\times 10^{23}$ 
& 3.4 $\times 10^{-16}$ m$^{5}$ & 2.0 $\times
10^{21}$ \\ 
\end{tabular}
\vspace{12pt}
\caption{ $1\sigma$ limits on $\alpha_{n} r_{0}^{n - 1}$ and $\Lambda_{n}$ in
Eq.~(\protect\ref{V_n}) from Spero, {\em et al.} [5] and Mitrofanov
and Ponomareva [13].}
\end{table}

\begin{figure}
\caption{Laboratory constraints on $g_{p}^{2}$ and $g_{n}^{2}$.  The region
in the $g_{p}^{2}$--$g_{n}^{2}$ plane above and to the right of each curve
is excluded at the $1\sigma$ level by the indicated experiment.  The gray
shading indicates the region excluded by the overlap of all present
laboratory experiments, and the remaining allowed  region is shown in
white.    The data are from  
Gundlach, {\em et al.} [2], Ramsey [16],  Ritter, {\em et al.} [15],
and Spero, {\em et al.} [5].   
The limit from Ritter, {\em et al.} on
$g_{n}^{2}$ is derived in Ref.~[1].
}
\end{figure}


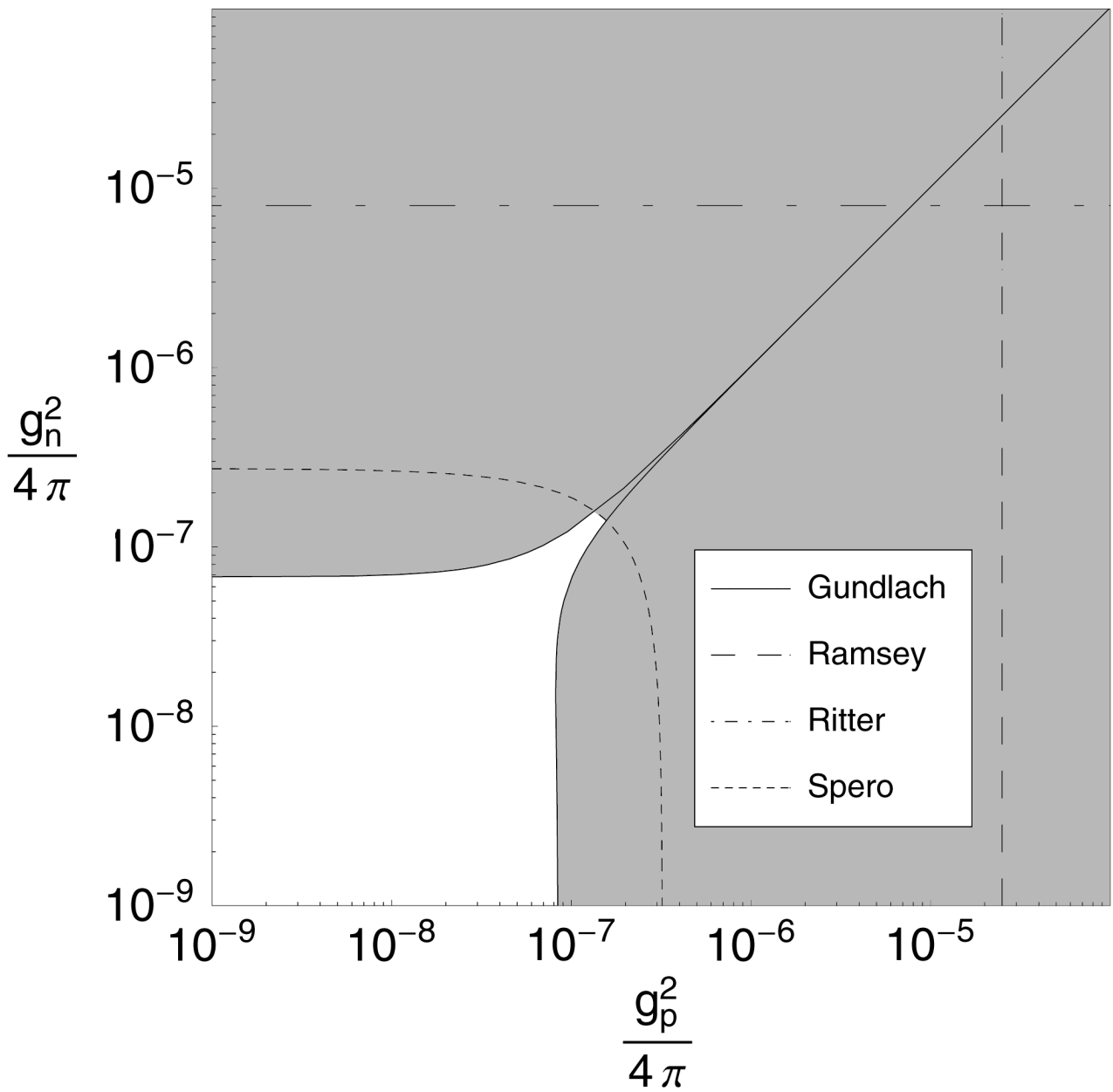
\end{document}